\documentclass[11pt]{article}
\usepackage[T2A]{fontenc}
\usepackage{booktabs}
\usepackage[utf8]{inputenc}
\usepackage{siunitx}
\usepackage{xcolor}
\usepackage{hyphenat}
\usepackage{geometry}
\usepackage{graphicx}
\usepackage{amsmath}
\usepackage{amssymb}
\usepackage{authblk}

\begin{document}

\title{Peculiar objects in the birthplaces of radio pulsars~--- stellar-mass black hole candidates}

\author{L.Chmyreva$^1$, G.Beskin$^{1,2}$}
\date{%
    $^1$Special astrophysical observatory of RAS,
  Nizhny Arkhyz, Russia\\%
    $^2$Kazan (Volga Region) Federal University, Kazan, 420008 Russia\\[2ex]%
    %\today
}

\maketitle
\begin{abstract}
We perform a search for stellar-mass black hole candidates in the spatial regions with increased probability of their occurrence, isolated based on the evolutionary scenarios for compact objects originating in disrupted binaries. We analyze the sources located in these regions with available spectral or photometric data, as well as measured proper motions and distances. Nine objects that exhibit characteristics corresponding to theoretical predictions for isolated black holes are marked for further study as black hole candidates.

\end{abstract}
 
\section{Introduction}

A key attribute of black holes (BH) is the presence of an event
horizon~--- a semipermeable membrane that confines the region of
space containing the black hole. In other words, in order to
identify the object in question as a BH, one must obtain data from
the regions located in direct proximity of the event horizon,
which is extremely difficult to do \cite{2005A&A...440..223B}.
Even in the outstanding experiment that imaged the central BH in
the M\,87 galaxy, such information came from a zone that is
located at a distance of 2.5 Schwarzschild radii from the event
horizon \cite{2019ApJ...875L...1E}. Indeed, both in galaxies and
in the accretion disks in X-ray binaries with BHs, the regions
that are in close vicinity of the event horizon are screened to a
significant degree by the surrounding accreting material. At the
same time, since the density of interstellar plasma is
sufficiently low \cite{2005A&A...440..223B}, the accretion rate
for isolated stellar-mass BHs is also low
\mbox{($10^{-6}$--$10^{-9}\dot{M}_{\rm{Edd}}$,} where
$\dot{M}_{\rm{Edd}}\approx10^{-8}\dfrac{M_\odot}{\rm
{yr}}\Big(\dfrac{M}{M_\odot}\Big)$, and $M$ is the BH mass),
allowing one to register photons originating in the immediate
proximity of the event horizon \cite{2008AdSpR..42..523B}.

The number of such isolated BHs, according to the latest evolutionary scenarios, is approximately $10^8$ for our Galaxy
\cite{2019ApJ...885....1W}. Isolated BHs usually undergo spherical accretion, since the velocity and density dispersions of the turbulent interstellar gas are small \cite{1952MNRAS.112..195B}.

The emission spectrum of the accretion flow for such a scenario was first obtained by Shvartsman \cite{1971SvA....15..377S} in 1971. He showed that in this case, synchrotron radiation of electrons is generated in the chaotic magnetic fields enhanced in the process of plasma inflow onto the BH. Most of the radiation originates in the areas close to the event horizon, at a distance of about $3$--$5\,r_g$, where $r_g$ is the gravitational radius. The luminosity of such a halo remains practically unchanged in a wide range of frequencies from $10^{14}$ to $10^{20}$~Hz; moreover, its spectrum has no lines \cite{1971SvA....15..377S, 1974Ap&SS..28...45B,
1975A&A....44...59M, 1982ApJ...255..654I}. These estimates laid the foundation for the critical MANIA experiment (Multichannel Analysis of Nanosecond Intensity Alterations) dedicated to the search for isolated stellar-mass BHs and based on detecting fast variations of the mentioned radiation. These luminosity variations manifest themselves as flares with typical time scales of $10^{-6}$--$10^{-3}$~s,
appearing when blobs of plasma, formed by the fragmentation of the accretion flow, approach the event horizon \cite{1971SvA....15..377S,
2008AdSpR..42..523B, 2005A&A...440..223B}.

Thus, for typical interstellar medium parameters, velocities, and masses, BHs would manifest themselves as objects with luminosities ranging from
$10^{28}$ to $10^{34}$~erg~s$^{-1}$, with continuous spectra in the interval from the infrared to the gamma region. Another common property would be light variations with amplitudes ranging from fractions of a percent to ten percent, with the duration of individual flares from $10^{-6}$ to $10^{-3}$~s. At the same time, these objects may be variable on time scales from months to years, which is due to the fact that a BH moves through an inhomogeneous interstellar medium (see
\cite{2008AdSpR..42..523B, 2005A&A...440..223B} and references therein).

Consequentially, a set of observational manifestations typical for BH candidates can be formulated as follows. They are Galactic sources that
 \begin{list}{}{
        \setlength\leftmargin{7mm} \setlength\topsep{2mm}
        \setlength\parsep{0mm} \setlength\itemsep{2mm} }
 \item[$1)$] radiate in a wide frequency range (from IR to gamma),
 \item[$2)$] have non-thermal continuous spectra with no lines,
 \item[$3)$] have visual magnitudes in the \mbox{$16^{\rm
{m}}$--$25^{\rm {m}}$} range (for typical distances of $100$--$300$~pc),
 \item[$4)$] demonstrate variations on long time scales (from hours to years),
 \item[$5)$] being located in our Galaxy, should have proper motions corresponding to linear velocities from approximately several to $100$~km\,s$^{-1}$, which for distances of up to about $400$~pc leads to angular displacements and/or parallaxes of up to $50$~mas~yr$^{-1}$ \cite{2018A&A...616A...1G}.
\end{list}

For further investigation, we selected objects that correspond to at least one or two of the noted criteria. In particular, some Galactic sources have properties similar to those that should be typical of isolated BHs. These are sources such as DC dwarfs (cool white dwarfs with a helium atmosphere and continuous spectra with no absorption lines), BL Lacertae objects (a blazar subtype characterized by non-thermal continuous spectra with no lines and fast variations), ROCOS objects (Radio Objects with a Continuous Optical Spectrum, also a blazar subtype), high-energy radiation sources unidentified in the visual range \cite{1989Afz....31..457S, 1989SvAL...15..145S,
2010AJ....139..390P}. As was already stated above, detecting short flares in the selected candidates is an important criterion for identifying them as BHs. Such candidate objects were isolated all over the northern sky within the framework of the ``MANIA'' experiment mentioned above and conducted since 1972 using a high temporal resolution photometric complex and the 6-meter telescope of the Special Astrophysical Observatory of RAS, after which searches were carried out for their rapid variations \cite{1977SoSAO..19....5S, 1989Afz....31..457S,
1989SvAL...15..145S, 2010AJ....139..390P}.

Among other strategies of searching for isolated BHs we should note studies of their emission in the radio and X-ray spectral regions. Modern X-ray observatories (NuStar,
Spektr--RG) have sufficient sensitivity to detect BHs, as was shown by, e.g., \cite{2002MNRAS.334..553A,2018MNRAS.477..791T}. The contribution of isolated BHs to the detected X-ray radiation can be comparable with the contribution from neutron stars; moreover, the hard spectral component is highly variable due to the emission from blobs of non-thermal electrons (a detailed description can be found in \cite{2005A&A...440..223B}). Isolated BHs can also be sources of unidentified gamma and radio emission in molecular clouds or cool neutral regions \cite{2012MNRAS.427..589B, 2005MNRAS.360L..30M}. Estimates of BH radio luminosity were obtained from a correlation between radio and X-ray luminosities derived for BHs in binary systems. The possibility of detecting BHs in the radio domain by future missions has been studied in detail by \cite{2019MNRAS.488.2099T}. Model computations
\cite{2013MNRAS.430.1538F,2021MNRAS.505.4036S} taking into account the distribution of masses and velocities, as well as the interstellar medium in the solar vicinity have shown that searching for BHs in the radio range is even more feasible, considering the sensitivity of the current and future projects, such as the Square Kilometer Array (SKA), which will be able to detect radio waves from nearby BHs. \cite{2019MNRAS.489.2038I} have shown the possibility of searching for isolated stellar-mass BHs in the central molecular zone of the Galaxy in the sub-millimeter and infrared spectral regions using space observatories. It thus follows from the aforementioned that detecting BHs is possible by observing the manifestations of their interaction with the surrounding medium. Another method of detecting isolated BHs is astrometric microlensing. Modeling has shown that this method can also be realized in observations (see, for example,
\cite{2016ApJ...830...41L}).

In this work we study objects in regions where the probability of locating a black hole is higher, using the above-mentioned characteristics to select the candidates. Such regions include zones of possible disruption of massive binary systems containing black holes and neutron stars (NS). We use the kinematic properties of relatively young pulsars to determine the positions of these zones. This process is described in Section~2. In Section~3 we select candidate BH sources in these regions. Section~4 presents a comparison of the observed properties of the candidates with the theoretical predictions for isolated BHs. The main results are outlined and discussed in Section~5.

\section{Determination of the Probable BH Localization Regions}
\subsection{NS and BH binary system disruption}

At least $70$\% of all stars are members of binary and multiple
systems \cite{2019ApJ...885....1W, 1967ARA&A...5...25B,
1991A&A...248..485D, 2003A&A...397..159H}. Therefore, one can
assume with a high probability that many of the now single
relativistic stellar-mass objects originated in binaries, either
as the final stage of evolution of the massive component
stars (NS are formed in the course of evolution of stars
with initial masses in the $10$--$25 M_\odot$ interval, whereas
BHs originate from collapsing stars with higher masses, \mbox{$M
\gtrsim 25 M_\odot$}), or as a product of merging of these stars
or the compact objects that formed from them. Modern gravitational wave
observations allow one to study mergers of the compact components
(see, e.g., \cite{2016ApJ...818L..22A}). Thus, modeling in
\cite{2018MNRAS.480.2704L} has shown that $7\times10^5$ binary
BHs have already merged in the Galaxy. In this work, we consider
the case where the system is disrupted and the former components
may be identified as members of the disrupted binary. The
evolutionary scenarios of BH and NS formation in binaries are well
described; see, for example, \cite{1998A&A...332..173P} and
\cite{2019ApJ...885....1W}. In binaries consisting of a BH\,+\,NS
pair, the BH is formed first in a supernova explosion in the end
of the evolution of the more massive star. In about $40$\% of the
cases the binary remains gravitationally bound. The second
supernova explosion, in which the NS is born, usually disrupts the
system \cite{1998ApJ...506..780B}. As a result of the explosion
asymmetry \cite{1996ApJ...456..738I, 1977ApJ...216..842H,
1969AZh....46..715S, 1987ApJ...321..780D} or a slingshot effect
\cite{1961BAN....15..265B, 1970ApJ...160L..91G}, neutron stars,
most of which manifest themselves as radio pulsars, attain high
velocities and become some of the fastest galactic objects: their
transverse velocities, determined from proper motions, reach
hundreds and even thousands of kilometers per second (see
\cite{2005MNRAS.360..974H} and references therein). An analysis
of pulsar kinematics allows one to trace their motion in the past
and, using their characteristic age estimates ($\tau_{\rm
ch}=P/2\dot P$, where $P$ is the pulsar period), determine the
probable location of their birth, accompanied by binary
disruption. This method was developed in our paper
\cite{2010AstL...36..116C}, where we conducted a search for pairs
of isolated pulsars that were previously members of binary
systems.

Since BH masses are several times higher than the masses of NSs, the BH will get a smaller velocity after system disruption and, therefore, will be located near the pulsar birthplace \cite{2019ApJ...885....1W}. In particular, for an approximately sevenfold mass ratio between these compact objects, the BH velocities turn out to be of the order of 5~km\,s$^{-1}$, which for a pulsar characteristic age of 500 thousand years gives about 2.6~pc for the BH displacement from the location of binary disruption ($0.75^{\circ}$ for a distance to the BH of 200~pc). The latter is smaller than the size of the region of the probable NS birth, which is determined by the kinematic properties of the NS (proper motion and distance uncertainties) and its age (the younger the NS, the more accurate its birth region coordinates and the smaller that region is). Thus, the a priori probability of a BH being located in these regions is enhanced, which allows us to narrow down significantly the search areas for these objects.

\subsection{Pulsar trajectories in the Galaxy}

We consider the motions of pulsars in a rectangular coordinate system whose origin is placed at the Galactic center. One of its axes is parallel to the direction of the Sun, the second axis is directed along the velocity of the local standard of rest (LSR), and the third axis is perpendicular to the Galactic plane and complements the first two to a right-handed vector triple \cite{1981gask.book.....M}. The law of pulsar motion ${\bf r}(t)$ in the Galactic gravitational potential $\varphi_G({\bf r})$ is the solution to the equation of motion
\begin{equation}
\ddot {\bf r} = -\nabla\varphi_G({\bf r})
\label{eqn:motion}
\end{equation}
with the initial conditions
\begin{equation}
{\bf r}_0 = {\bf r}(t = 0), {\bf V}_0 = {\bf V}(t = 0),
\label{eqn:initial}
\end{equation}
corresponding to the present epoch. The gravitational potential $\varphi_G({\bf r})$ has the following form
\cite{1987AJ.....94..666C, 1989MNRAS.239..571K}:
\begin{equation}
\varphi_G(r,z)=-\frac{GM_{dh}}{\sqrt{(a_G+\sum_{i=1}^3\beta_i\sqrt{z^2+h_i^2})^2+b_{dh}^2+r^2}}-\frac{GM_b}{\sqrt{b_b^2+r^2}}-\frac{GM_n}{\sqrt{b_n^2+r^2}}.
\label{eqn:phi}
\end{equation}

This is a three-component axisymmetric function that includes the contributions from the disk and halo, bulge, and nucleus, labeled by the indices dh, $b$
and $n$ correspondingly. Here $r$ is the distance from the center of the Galaxy, $z$ is the distance from the Galactic plane, and $M,\beta_i, h_i, a_G, b$ are constants (see
\cite{2010AstL...36..116C}).
Since the Galactic potential is not spherically symmetric, the solution to equation~(1) is generally obtained numerically
\cite{1987AJ.....94..666C, 1989MNRAS.239..571K}.
Vectors ${\bf r}_0$ and ${\bf V}_0$ are determined from the pulsar distance and velocity data obtained in observations. Their measurement accuracy determines the scatter of the pulsar trajectories and, as a result, the size of the probable BH localization region. For the radius vector ${\bf r}_0$ of a pulsar with galactic coordinates $l$ and $b$ and a heliocentric distance $d$, we have
\begin{equation}
{\bf r}_0 = d \cdot ({\bf i}\cdot\cos b \cos l+{\bf j}\cdot\cos b \sin l+{\bf k}\cdot\sin b) + {\bf r}_{\odot},
\label{r0}
\end{equation}

where ${\bf i},{\bf j},{\bf k}$ are unit vectors in the Cartesian coordinate system mentioned above and \mbox{${\bf r}_{\odot}=-{\bf
i}\cdot(8.5$ kpc)} is the radius vector of the Sun. The pulsar velocity vector ${\bf
V}_0$ is determined by its proper motion components $\mu''_l$ and $\mu''_b$,
distance $d$, radial velocity $V_{r}$, and the solar velocity vector:
\begin{equation}
{\bf V}_0 = {\bf V}_{r} + {\bf V}_t + \dot {\bf r}_{\odot}.
\label{V0}
\end{equation}

Here $\dot{\bf r}_{\odot}\!=\!{\bf V}_{\odot,\,\rm{rot}}\!+\!{\bf
V}_{\odot,\,\rm{LSR}}$,
where ${\bf V}_{\odot,\,\rm{rot}} = {\bf j} V_{\odot,\,\rm{rot}}$ is the LSR
Galactic plane rotation velocity and ${\bf V}_{\odot,\,\rm{LSR}}$ is the solar velocity relative to the LSR \cite{1981gask.book.....M}. The pulsar transverse velocity is determined from its proper motion and distance: $V_{t} = 4.74 d \sqrt{ (\mu''_l \cos b)^2 + {\mu''_b}^2 },$ where $\mu''_l$ and
$\mu''_b$ are expressed in [mas yr$^{-1}$], and $d$ in kiloparsecs. The radial velocity of a pulsar ${\bf V}_{r}={\bf V}_{r,\,\rm{rot}}+{\bf V}_{r,\,p}$ consists of a secular component ${\bf V}_{r,\,\rm{rot}}$, attributable to the Galactic rotation of the pulsar LSR, and peculiar velocity ${\bf V}_{r,\,p}$.

Pulsar proper motion components ($\mu''_l$, $\mu''_b$) or their upper limits are determined from observations. Distances are obtained from parallaxes $\pi$, with an average relative precision of about 20\% \cite{2005yCat.7245....0M}, or from dispersion measures, with an accuracy of about 30\% \cite{2002astro.ph..7156C}. In the latter case, they are model-dependent.

The main problem when studying pulsar kinematics is determining their radial velocities, which are projections of their total velocities onto the line of sight. They can be described in two ways, with the initial assumption of their isotropic distribution. Specifically, using a one-component function derived in
\cite{2005MNRAS.360..974H} from an analysis of proper motions for a sample of 233 pulsars, or a two-component distribution, derived in
\cite{2002ApJ...568..289A} by means of modeling and comparing with radio survey data. At the same time, as was shown in \cite{2006ApJ...643..332F}, all models are consistent with observations and it is hard to choose between these two options. Thus, the choice of a particular distribution is not so important~--- here we use the distribution from
\cite{2002ApJ...568..289A} for the peculiar component of the pulsar radial velocity.

Following \cite{2004ApJ...610..402V}, we assume that the pulsar proper motions are distributed according to normal laws $N(\mu''_l,
\sigma^2_{\mu''_l})$ and $N(\mu''_b, \sigma^2_{\mu''_b})$, where the means and variances are equal to the measured angular velocities and the squares of their errors correspondingly. The distribution of distances to the objects has a similar form $N(d, \sigma^2_d$). We then derive from formulas (\ref{r0}) and (\ref{V0}) the distributions $p_r({\bf
r}_0)$ and $p_v({\bf V}_0)$ for ${\bf r}_0$ and ${\bf V}_0$. Based on the latter, we use the Monte Carlo method to simulate a series of pulsar trajectories and determine their localization regions corresponding to the epoch $t_0-\tau_{\rm ch}$, where $t_0$ is the current epoch and $\tau_{\rm ch}$ is the characteristic age. This method is discussed in detail in our paper \cite{2010AstL...36..116C}, where it was tested on several pulsar pairs in order to determine their possible past connections in binary systems.

\subsection{Determining the pulsar birthplaces}

Based on the assumption that the pulsars were members of binary
systems, the sizes of their disruption regions should be close to
the uncertainties of gamma- and X-ray sources
\cite{2000A&AS..143...33B}; considering the errors of the
initial pulsar coordinates found from their trajectories, we
estimated the upper limits for the ages of the considered objects.
With the coordinate uncertainties of gamma- and X-ray sources reaching
several degrees, these limits are close to one million years.
Based on this assumption, we selected 16 isolated radio pulsars
from the ATNF
database\footnote{http://www.atnf.csiro.au/research/pulsar/psrcat/expert.html},
with measured proper motions and parallaxes.

\begin{figure}[ht!]
   \centering
    \includegraphics[width=0.8\textwidth]{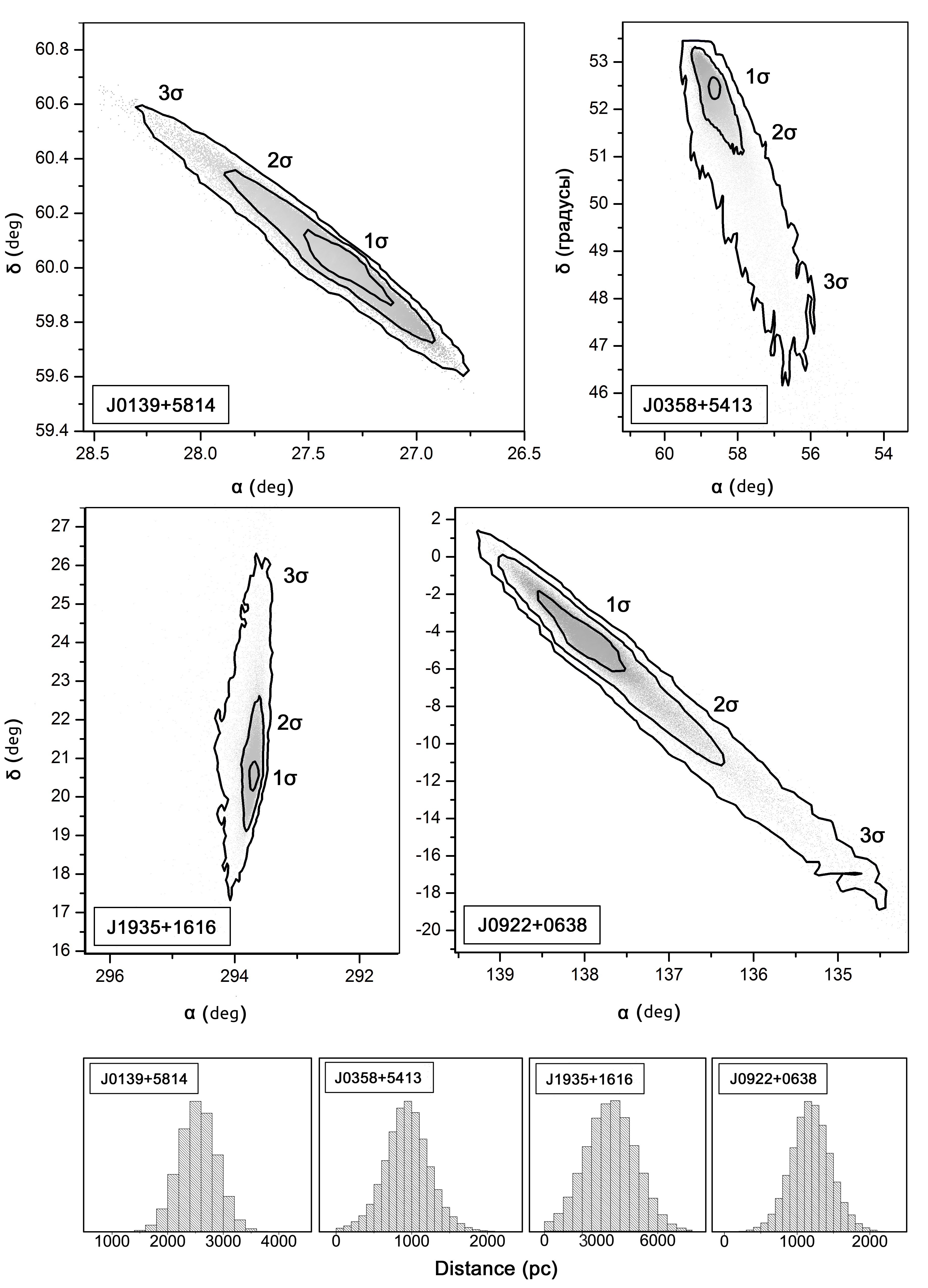}
\caption{\small{Top: birth locations of pulsars J\,0139+5814, J\,0358+5413, J\,1935+1616 and J\,0922+0638 projected onto the celestial sphere, where a search for BH candidates was conducted. The coordinates are given in degrees. The grey dots represent the endpoints of the simulated pulsar trajectories corresponding to their characteristic ages. The black contours show the 1$\sigma$, 2$\sigma$ and 3$\sigma$ areas. Bottom: histograms of the trajectory endpoint distributions by distance from Earth in parsecs (radial localization of the regions).}}
\label{sigmas figure}
\end{figure}

Modeling the
distribution of pulsar spatial velocities and solving
equation~(\ref{eqn:motion}) of their motion in the Galactic
potential, we simulated 100,000 trajectories for each pulsar
(where their initial coordinates were determined as described in
the previous section), which were traced into the past up to the
moment of time corresponding to the current characteristic age
estimate. The distribution of the endpoints of these trajectories
determined the spatial region where the suggested binary
disruption occurred and where a BH is possibly located.
Fig.~\ref{sigmas figure} shows the localization regions with the
contours corresponding to 1$\sigma$, 2$\sigma$ and 3$\sigma$-levels for
the four youngest pulsars out of 16 (J\,0139+5814, J\,0922+0638,
J\,0358+5413 and J\,1935+1616), whose parameters are presented in
Table~\ref{tab:psr parameters}. The areas of the probable binary
disruption regions range from one to 16 square degrees. For the
remaining pulsars, these areas are significantly larger.

\begin{table}[ht!]\begin{center}
\begin{tabular}{llllll}\toprule\toprule
  Pulsar & Position & Proper motion & Distance & Age \\ 
          & J2000  ($\alpha$, $\delta$) & $\mu''_{\alpha}$, $\mu''_{\delta}$ (mas yr$^{-1}$) & (pc) & (years)  \\ 
  \midrule
  
  \scriptsize{J0139+5814} & \scriptsize{\, 01 39 19.7401}& \scriptsize{-19.11(0.07)} & \scriptsize{2600(300)}  &  \scriptsize{403\,000}\\
       &\scriptsize{+58 14 31.819}& \scriptsize{-16.60(0.07)} &  &   \\
  
  \scriptsize{J0922+0638}& \scriptsize{\, 09 22 14.022} &  \scriptsize{\,18.8(0.9)} & \scriptsize{1100(200)}  &  \scriptsize{497\,000}\\
  & \scriptsize{+06 38 23.30}& \scriptsize{\,86.4(0.7)} &   &   \\
  
  \scriptsize{J0358+5413}&\scriptsize{\, 03 58 53.71650} &  \scriptsize{\,9.20(0.18)} & \scriptsize{1000(200)}  & \scriptsize{564\,000}  \\
    &\scriptsize{+54 13 13.7273} & \scriptsize{\,8.17(0.39)} &   &  \\

\scriptsize{J1935+1616} &\scriptsize{\, 19 35 47.8259} & \scriptsize{\,1.13(0.13)} & \scriptsize{3700(1300)}  & \scriptsize{947\,000} \\
  & \scriptsize{+16 16 39.986}& \scriptsize{-16.09(0.15)} &   &   \\ \bottomrule
  
\end{tabular}\caption{\small{Parameters of pulsars with minimal ages, whose birthplaces were used to search for BH candidates} }\label{tab:psr parameters}
\end{center}\end{table}

\clearpage

\section{Selection of BH Candidates Among Optical Sources}

\subsection{Peculiar optical objects in the probable pulsar birthplaces}

At the first stage, according to the criteria formulated above, we used the Aladin \cite{2000A&AS..143...33B} database to select in the probable pulsar birthplaces, shown by the 3$\sigma$ contours in Fig.~\ref{sigmas
figure}, 59~white dwarfs and blue objects (with color indices $B-V<0$)
\cite{1991BSAO...31...33B}. Objects precisely of this type, with no spectral lines, were targeted in the ``MANIA'' experiment.

At the second stage, these same regions were used to select optical sources that fall within the overlapping coordinate uncertainty ellipses of gamma, X-ray, and radio sources. To this end, we selected 5~(non GRB) gamma-ray sources, 258~X-ray sources and 1896~radio sources from
ROSAT\footnote{https://heasarc.gsfc.nasa.gov/docs/rosat/rosat.html},
FERMI\footnote{https://fermi.gsfc.nasa.gov/},
XMM--Newton\footnote{https://www.cosmos.esa.int/web/xmm-newton},
FIRST\footnote{http://sundog.stsci.edu/} databases. In particular, the birthplace of pulsar J\,0139+5814 has 3~X-ray sources and 19~radio sources, the J\,0922+0638 birthplace~--- 87~X-ray,
257~radio, and 2~gamma-ray sources, the J\,0358+5413 birthplace~--- 98~X-ray and
885~radio sources, and the J\,1935+1616 birthplace~--- 3~gamma, 70~X-ray and
737~radio sources. Cross identification of these sources within the positional error ellipses, which are minimal for the X-ray sources and amount to about $25''$ for a 1$\sigma$-level error, lead to 57~matches between X-ray and radio sources. Forty five of these are quasars, stars, or galaxies \cite{2000A&AS..143....9W}. Twelve regions were finally selected with sizes of about $10''$--$20''$, where the difference between the coordinates of the radio and X-ray sources does not exceed their positional errors. These areas host
35~optical objects from the SDSS\footnote{https://www.sdss.org/},
DSS\footnote{https://irsa.ipac.caltech.edu/data/DSS/},
CDS\footnote{http://cdsportal.u-strasbg.fr/} databases. Thus, a sample was formed for further study, consisting of 94~optical objects,
59~of which were selected by color, and 35~--- by the coordinate uncertainties of the X-ray sources.

\subsection{Analysis of the properties of the selected objects}

The following properties of these 94 sample objects were analyzed:
\begin{list}{}{
        \setlength\leftmargin{7mm} \setlength\topsep{2mm}
        \setlength\parsep{0mm} \setlength\itemsep{2mm} }
\item[$1.$]\textit{Morphology.} In the course of analysis, five unidentified extended sources were discovered that were present in only one frame and which, in all likelihood, are artifacts; we excluded them from further consideration.

\item[$2.$]\textit{Kinematics.} Six objects do not have proper motions (at a level of $\mu''<$ 0.6, 0.61, 3.0, 1.3, 0.88 and 1.9 mas\,yr$^{-1}$)
\cite{2018A&A...616A...1G}. Since the detection of proper motion directly points to the Galactic localization of an object, these six candidates were excluded from further analysis.

\item[$3.$]\textit{Distances.} GAIA parallax data leads to distance estimates for 10~objects of more than 2525~pc, 2389~pc and 9415~pc, which are the upper limits for model birthplaces of pulsars
J\,0922+0638, J\,0358+5413 and J\,1935+1616. If these objects were BHs, their magnitudes would be fainter than $25^{\rm m}$ \cite{2005A&A...440..223B}; for this reason we excluded them from the sample.

\item[$4.$]\textit{Spectral characteristics.} Four objects have line spectra (1~star, 3~quasars) \cite{2000A&AS..143....9W} and have also been excluded from the list.

\item[$5.$]\textit{Photometry.} Medium-band photometric data are available for 57~objects (up to $ 10$ bands \cite{2000A&AS..143....9W}). The VOSpec software was used to fit these data with Planck curves. The fitting accuracy was sufficient (about 10\%) to identify the spectra of these objects as thermal; they were excluded from future consideration.
\end{list}

%GAIA\footnote{https://www.cosmos.esa.int/web/gaia}

The final sample consisted of 12~objects. Three of them do not have distance data available and further analysis was not carried out for them within the framework of this paper. The nine remaining candidates are mainly faint sources
$19^{\rm m}$--$21^{\rm m}$, located at distances up to about 500~pc. A comparison of their luminosities with the theoretical expectations for BHs is presented in the next section.

\begin{figure}[ht!]
   \centering
    \includegraphics[width=1.0\textwidth]{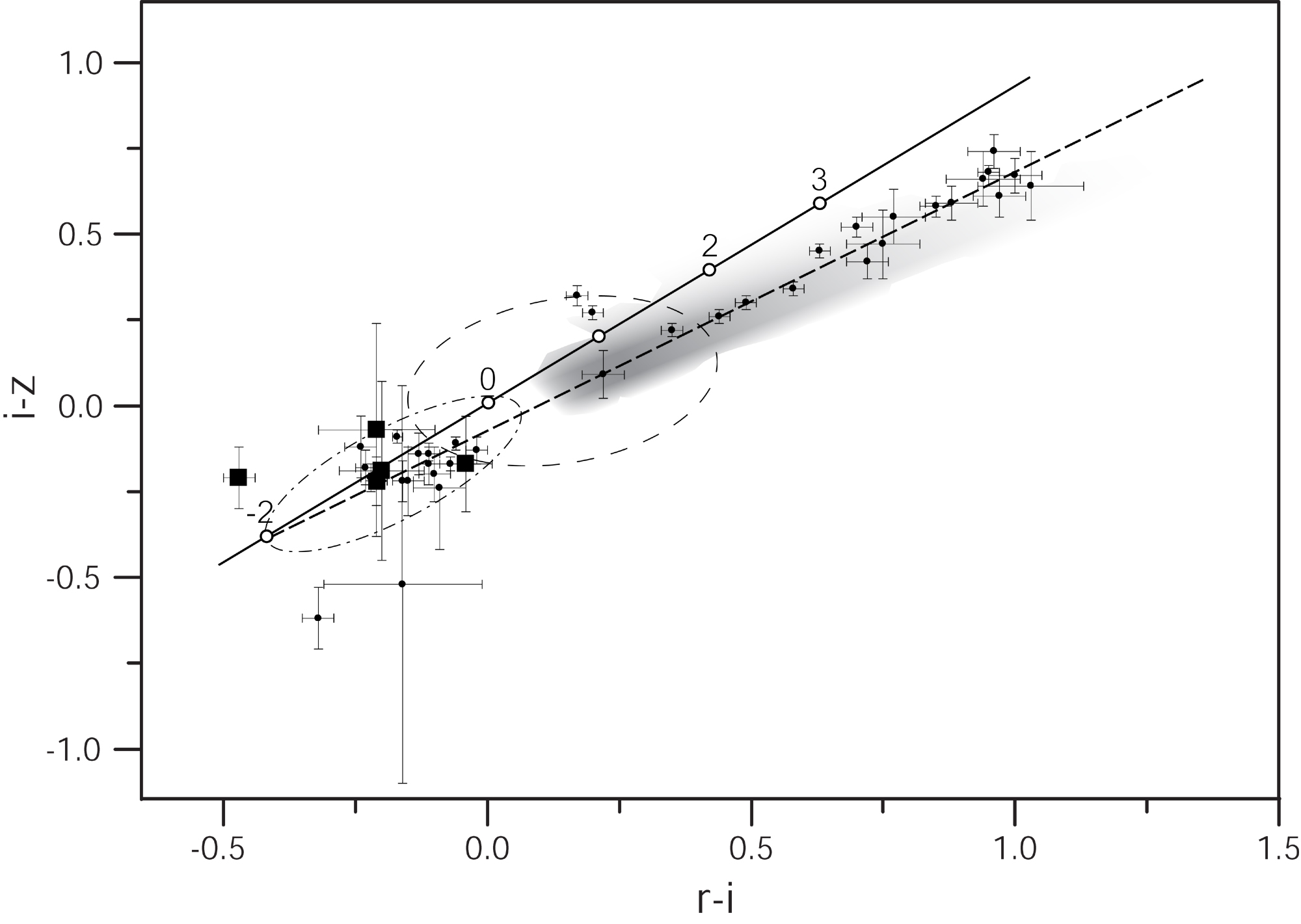}
\caption{\small{``Color--color'' diagram for the sample objects with SDSS measurements available. The shaded area shows the stellar locus determined from background stars in the considered areas. The dashed ellipse corresponds to the location of quasars in the diagram, and the dashed-and-dotted ellipse~---to that of white dwarfs. The solid and dashed diagonal lines show the positions of sources with power and blackbody spectra correspondingly. SDSS data are available for five of the nine sources of the final sample (see below)~--- they are marked by squares.}}
\label{color-color figure}
\end{figure}

SDSS colors are available for 48~objects out of the 94~mentioned above; they are presented in Fig.~\ref{color-color figure}. Thus, the objects in this sample are clearly divided into two categories: 20~of them fall within the category of normal stars (they have Planck spectra, see above), and 28~fall into the white dwarf and quasar domain \cite{2005AJ....129.2542C, 2011MNRAS.417.1210G}. Note that the $(i-z)-(r-i)$ colors for six of the 12~final candidates are located in the latter, however, it does not seem possible to ascertain whether their spectra are thermal or non-thermal, given the accuracy of the color determination.

\section{Comparison of the Observed Properties of the Selected Objects with their Theoretical Estimates}
\subsection{Determining the acceptable masses and velocities for hypothetical black holes that manifest themselves as the selected candidates}

Since the theoretical concepts about the luminosity of halos around isolated BHs are based on spherical accretion models
\cite{1952MNRAS.112..195B,1971SvA....15..377S}, its estimates, on the one hand, are determined by the local interstellar medium parameters for the object (specifically, density and temperature), and on the other hand, by the mass and velocity of the BH itself. Most isolated stellar-mass BHs originate in binaries \cite{2019ApJ...885....1W}. This circumstance determines their masses and velocities after system disruption. This being the case, we estimate accordingly the theoretical magnitudes for the nine candidates and compare them with those observed.

Based on the classic accretion mechanism concepts
\cite{1944MNRAS.104..273B,1971SvA....15..377S}, the formula for the theoretical luminosity of an isolated BH with spherical accretion, which was used to estimate the luminosities of the nine candidates, can be written as
\cite{2005A&A...440..223B}:

\begin{equation}
L=9.6\times10^{33}M^{3}_{10}n^2(V^2+c^2_s)^{-3}_{16}\;\;\text{erg/s},
\label{L}
\end{equation}
where $M_{10}$ is the BH mass in units of 10$M_\odot$, $n$ is the medium density in units of cm$^{-3}$, and $V$ and $c_s$ are the total spatial BH velocity and the sound speed normalized to 16~km\,s$^{-1}$. The optical
$V$-band luminosity amounts to about 20\% of the total luminosity \cite{2005A&A...440..223B}. This expression, accurate up to a numerical coefficient, was first derived by
\cite{1971SvA....15..377S} in the assumption of maximal Bondi-Hoyle accretion rate
(\cite{1944MNRAS.104..273B}) (here the accretion efficiency coefficient $\alpha$
lies in the range of~$1$--$2$). The conclusion about the low efficiency of accretion itself in numerous works (see, for example,
\cite{2003ApJ...594..936P,2013MNRAS.430.1538F,2021MNRAS.505.4036S}) is made based on the absence of radio and X-ray range manifestations of isolated accreting neutron star and black hole populations. We believe that this conclusion is attributable to the rather specific assumptions about the conversion mechanisms of the accretion flow energy to radiation, associated with its density, magnetic field strength and variations, degree of isotropy, etc. At the same time, for low accretion rates and its spherical nature we can assume the accretion rate to be maximal and obtain the characteristics of its emission for physically justified concepts about the plasma flow parameter variations. All of the above has been carried out in \cite{2005A&A...440..223B}, and our analysis is based on that work. Note that the same expressions were derived for luminosities in \cite{1982ApJ...255..654I} and
\cite{1974Ap&SS..28...45B}, but with numerical factors several time smaller, which is due to the fact that in \cite{2005A&A...440..223B} we used a detailed electron heating model that takes into account the magnetic field influence on this process.

Using standard expressions for luminosity, magnitude and distance modulus, we derive for a BH the relation between mass and velocity for the interstellar medium parameters mentioned above and visual magnitude $m$ of an object at distance $D_{10}$ in units of 10~pc:

\begin{equation}
M=55.44\times10^{-(2/15)m}D^{2/3}_{10}n^{-2/3} (V^2+c^2_s)_{16}.
\label{MV}
\end{equation}

The quantities in (\ref{MV}), with account of their accuracies, determine a region on the $MV$ plane where the theoretical magnitude of a candidate agrees with the observed magnitude. In other words, the distance, density, and magnitude errors set the width of this region.

To determine $n$ we used empirical dependences $E_{g-r}(\mu)$ between the reddening $E_{g-r}$ and distance modulus $\mu$ for different directions, obtained from the data of the 3D Galactic dust map \cite{2019ApJ...887...93G}. Since $N_H=6.86\times 10^{21}
E_{B-V}$ \cite{2009MNRAS.400.2050G} and $E_{B-V}=0.884 E_{g-r}$
\cite{2019ApJ...887...93G}, we have for the hydrogen column density
$N_H=f(\mu)=6.06\times 10^{21} E_{g-r}(\mu)$. Differentiating this function and taking into account the relation between $D$ and distance modulus $\mu=5\lg(\frac{D}{10})$, we derive the local medium density $n$ in the vicinity of the object: 

\begin{equation}
n=\frac{dN_H}{dD}=6.06\times 10^{21}\frac{ dE_{g-r}(\mu)}{dD(\mu)}=1.3\times10^{21}\times10^{-\mu/5}\frac{dE_{g-r}(\mu)}{d\mu}.
\label{n}
\end{equation}

\begin{figure}[ht!]
   \centering
    \includegraphics[width=1.0\textwidth]{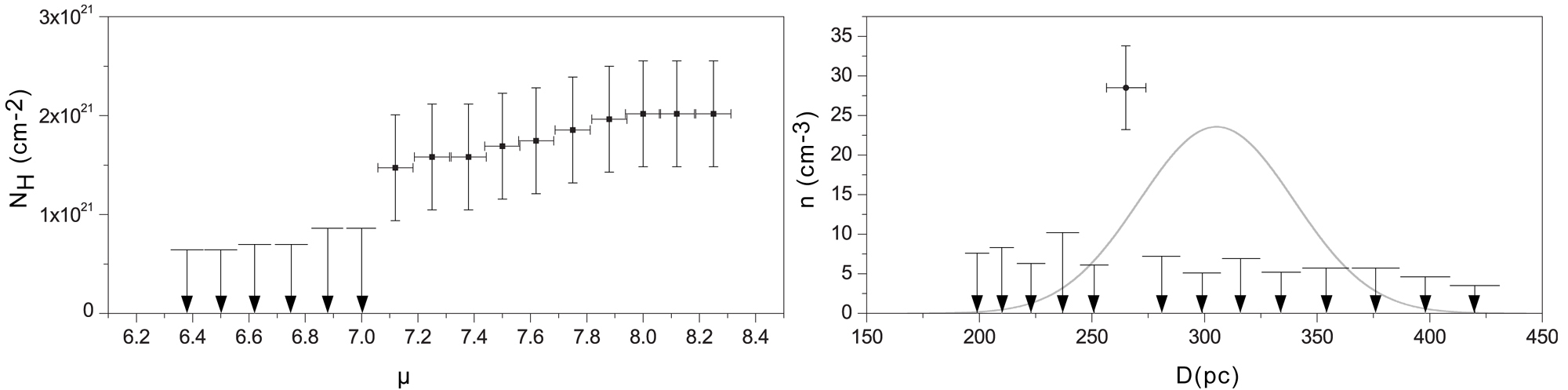}
\caption{\small{Left: sample $N_{H}$ column density distribution as a function of distance modulus $\mu$ for one of the candidates
(J\,193415.78+190004.2). The errors correspond to the $1\sigma$-level according to the distribution \cite{2019ApJ...887...93G}. The upper limits (for a $99$\% confidence level) are shown by arrows. Right: interstellar medium density $n$ estimate as a function of distance $D$. The Gaussian shows the distribution of the object distance estimates. }}
\label{fig:nd}
\end{figure}

Figure~\ref{fig:nd} illustrates this procedure for one sample object
(J\,193415.78+190004.2). The left panel shows the column density dependence on distance modulus from \cite{2019ApJ...887...93G} in the direction towards this object. The right panel shows the final dependence of density on its distance. The values of $n$ measured for each object are presented in Table~2, with the errors given in parentheses. Since the distance uncertainties for the objects are rather large, it is impossible to obtain an exact density value in their vicinity. As an estimate, we used a weighted-average value for the range of acceptable distances, determined by their measurement accuracy.
For this estimate, we have $\overline{n}\in(\overline{n}_{min},\overline{n}_{max})$ with $\overline{n}_{min}=\frac{\sum w n}{\sum w}$ and $\overline{n}_{max}=\frac{\sum w n+\sum w n_{0}}{\sum w}$ where
$n$ are the individual density values, $w$ are their probabilities, and $n_0$~is the upper density limit estimate.

Since each object has only the transverse component of its total velocity measured, this quantity represents its minimal value and limits the domain of acceptable velocities $V$ on the $MV$ plane. Additionally, $V>V_{\rm
tr}=4.74\mu{''}D$, where $V_{\rm tr}$~is the transverse velocity obtained from the observed proper motion
$\mu''=\sqrt{\mu_{\alpha}^{''2}+\mu_{\delta}^{''2}}$ (where $\mu''_{\alpha}$ and
$\mu''_{\delta}$~are the components along the $\alpha$ and $\delta$ coordinates).

Finally, we estimate the last parameter included in expression (\ref{L})~--- the local sound speed $c_s$. This quantity was determined using the standard formula
$c_s=\sqrt{\frac{\gamma kT}{m_p}}$, where $m_p$ is the proton mass, and the temperature
$T$ was estimated from the empirical dependence $T(n)$ from \cite{1981sss..book..265B} and its range was determined by the density interval $\overline n_{\rm
min},\overline n_{\rm max}$. The set of parameters described above is given for each object in Table~2.

\begin{table} \begin{center}
\begin{tabular}{llllll}\toprule\toprule
  Object & $\mu''_{\alpha}$, $\mu''_{\delta}$& $D$ & $\overline n$ & $V_{tr}$ & $m$\\ 
    & mas yr$^{-1}$ & pc & \SI{}{\centi\meter\tothe{-3}} & \SI{}{\kilo\meter\per\second} &  \\ 
  \midrule
  %34
  \scriptsize{J035738.16+525934.4} & \scriptsize{2.56$\pm0.88$, 6.58$\pm0.69$} & \scriptsize{392$\pm70$}  & \scriptsize{$\le4.0$} & \scriptsize{13$\pm3$}& \scriptsize{19.3$\pm0.4$} \\

    \hline
  %3a
  \scriptsize{J035757.63+525928.7} &  \scriptsize{-12.5$\pm1.5$, 2.6$\pm1.1$} & \scriptsize{377$\pm100$}  &  \scriptsize{$\le2.9$} & \scriptsize{23$\pm7$} & \scriptsize{19.6$\pm0.6$} \\

    \hline
  %3i
 \scriptsize{J035717.10+511525.4} &  \scriptsize{0.88$\pm0.51$, -1.8$\pm0.35$} & \scriptsize{571$\pm70$}  & \scriptsize{$\le2.8$} & \scriptsize{5$\pm1$} & \scriptsize{18.6$\pm0.3$}  \\

     \hline
  %3j
  \scriptsize{J035239.08+513344.1} & \scriptsize{18.9$\pm0.5$, -26.1$\pm0.4$} & \scriptsize{210$\pm10$}  & \scriptsize{7.2$\pm5.4$} & \scriptsize{32$\pm2$} & \scriptsize{18.7$\pm0.1$}  \\ 

     \hline
  %4d
  \scriptsize{J193559.98+205305.7} &  \scriptsize{7.4$\pm0.7$, 2.7$\pm0.8$} & \scriptsize{466$\pm120$}  & \scriptsize{4.8$\pm1.7$} & \scriptsize{17$\pm5$} & \scriptsize{19.7$\pm0.6$}  \\

     \hline
  %4f
\scriptsize{J193433.81+203117.1} &  \scriptsize{15.7$\pm0.3$, -36.1$\pm0.4$} & \scriptsize{290$\pm20$}  & \scriptsize{6.4$\pm3.0$} & \scriptsize{54$\pm4$} & \scriptsize{18.8$\pm0.2$}  \\

   \hline
    %4q
\scriptsize{J193415.78+190004.2} &  \scriptsize{0.5$\pm1.2$, -41.7$\pm1.7$} & \scriptsize{305$\pm100$}  & \scriptsize{3.5$\pm2.4$} & \scriptsize{60$\pm20$} & \scriptsize{20.5$\pm0.8$}  \\

   \hline
    %3n
\scriptsize{J034803.12+505358.7} &  \scriptsize{-1.97$\pm1.13$, -7.59$\pm0.92$} & \scriptsize{448$\pm200$}  & \scriptsize{2.4$\pm1.4$} & \scriptsize{17$\pm8$} & \scriptsize{20.5$\pm1.2$}  \\

  \hline
    %2с
\scriptsize{J090946.77--062229.8} &  \scriptsize{5.7$\pm0.9$, -21.4$\pm0.8$} & \scriptsize{507$\pm100$}  & \scriptsize{$\le0.9$} & \scriptsize{53$\pm10$} & \scriptsize{18.6$\pm0.5$}  \\ 
\bottomrule
  
\end{tabular}\caption{\scriptsize{Parameters and their errors for 9 selected candidates. Proper motions $\mu''_{\alpha}$, $\mu''_{\delta}$,
distances $D$, weighted-average densities $\overline n$, transverse velocities
$V_{\rm tr}$ and visual magnitudes $m$ are given.}}\label{tab:source parameters}
\end{center}\end{table}

\begin{figure}[ht!]
   \centering
    \includegraphics[width=1.0\textwidth]{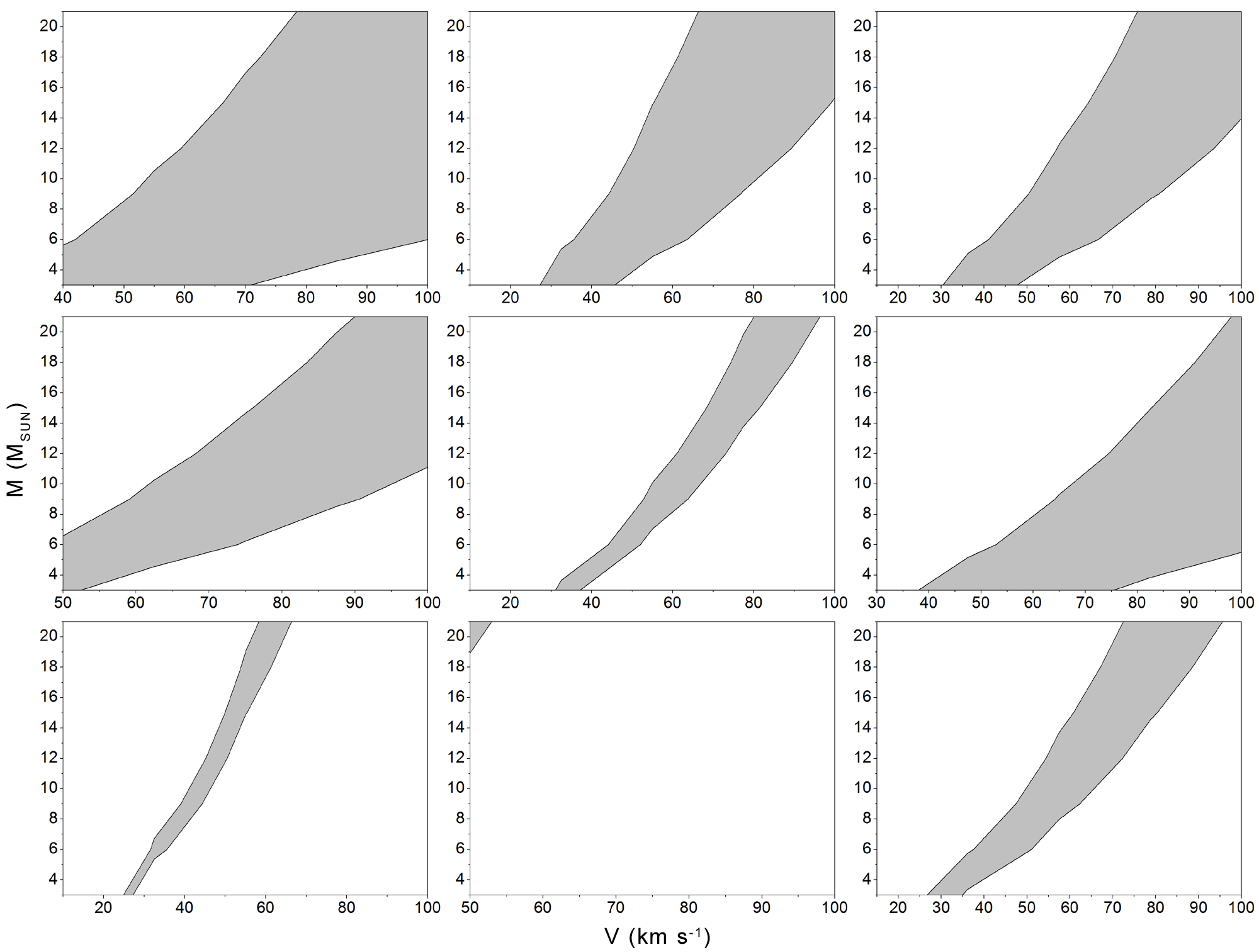}
\caption{\small{Areas of the probable velocity and mass localization for 9~selected objects. }}
\label{fig:all-ave}
\end{figure}

Fig.~\ref{fig:all-ave} shows the areas of probable velocity and mass localizations for the 9 selected objects. Their boundaries were determined according to the relation
(\ref{L}), using the acceptable intervals of these characteristics from Table~2.

\subsection{Estimating the probability that the selected candidates are stellar-mass black holes}

The combined regions of paired velocity and mass values for hypothetical BHs were thus obtained for 9~candidates, corresponding to their observed characteristics (see
Fig.~\ref{fig:all-ave}). We must now estimate the probability that these values agree with the existing models of birth and evolution of isolated BHs. As was already mentioned above, most BHs originate in disrupted binary systems. The previous evolutionary process~--- the kick obtained during the collapse and/or binary disruption~--- will influence the formation of these disrupted binary components. Modern evolutionary scenarios for isolated BHs are discussed in detail in \cite{2019ApJ...885....1W} and here we use the results of that work. The authors show that the greatest number of isolated BHs from disrupted binaries are formed in the case of solar metallicity. The standard evolution model for the Galactic disk population has solar metallicity, an isotropic initial kick distribution, and initial velocities from the Maxwellian distribution with $\sigma=265$~km\,s$^{-1}$ for compact objects \cite{2005MNRAS.360..974H}. Such a model yields a two-peak distribution for the velocities: slow BHs with a peak at low velocities (formed without an initial kick due to massive fallback or in the process of direct collapse) and fast BHs, with a broader wing at high velocities (BHs without sufficient fallback).

\begin{figure}[ht!]
   \centering
    \includegraphics[width=0.7\textwidth]{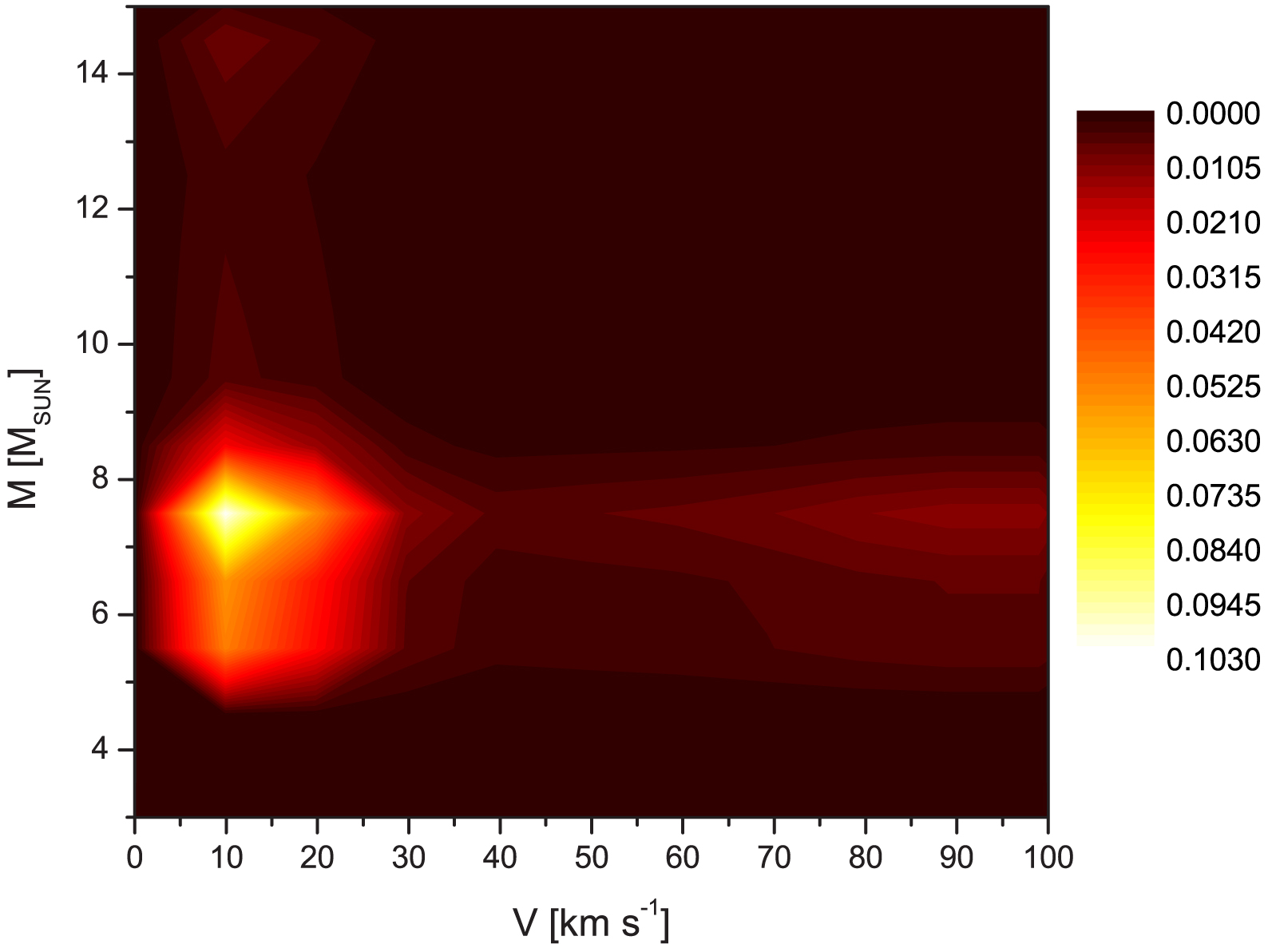}
\caption{\small{Two-dimensional distribution of the most probable masses $M$ and velocities $V$ for isolated BHs born in disrupted binaries. Probabilities are shown in shades of color.}}
\label{fig:w-prob}
\end{figure}

The mass distribution has the main peak at about 7--8~$M_\odot$ (BHs formed in the collapse of stars of approximately 20--35~$M_\odot$), a secondary, smaller peak at approximately $15~M_\odot$ (BHs originating from the most massive stars that lose some of their mass in the form of stellar wind), and a third, smallest peak near $22~M_\odot$ (a result of binary interaction).

Converting these independent distributions to probability distributions and multiplying them, we derive a two-dimensional space showing the regions of the most probable $M$ and $V$ values for isolated BHs that originated from binary systems~--- it is shown in Fig.~\ref{fig:w-prob}. As is evident from the figure, a distinct peak stands out at $V\sim10$~km\,s$^{-1}$ for a mass of about $7.5~M_{\odot}$, with less conspicuous ``wings'' at the same values and with practically zero probabilities in the remainder of the field.

\begin{figure}[ht!]
 \centering
    \includegraphics[width=1.0\textwidth]{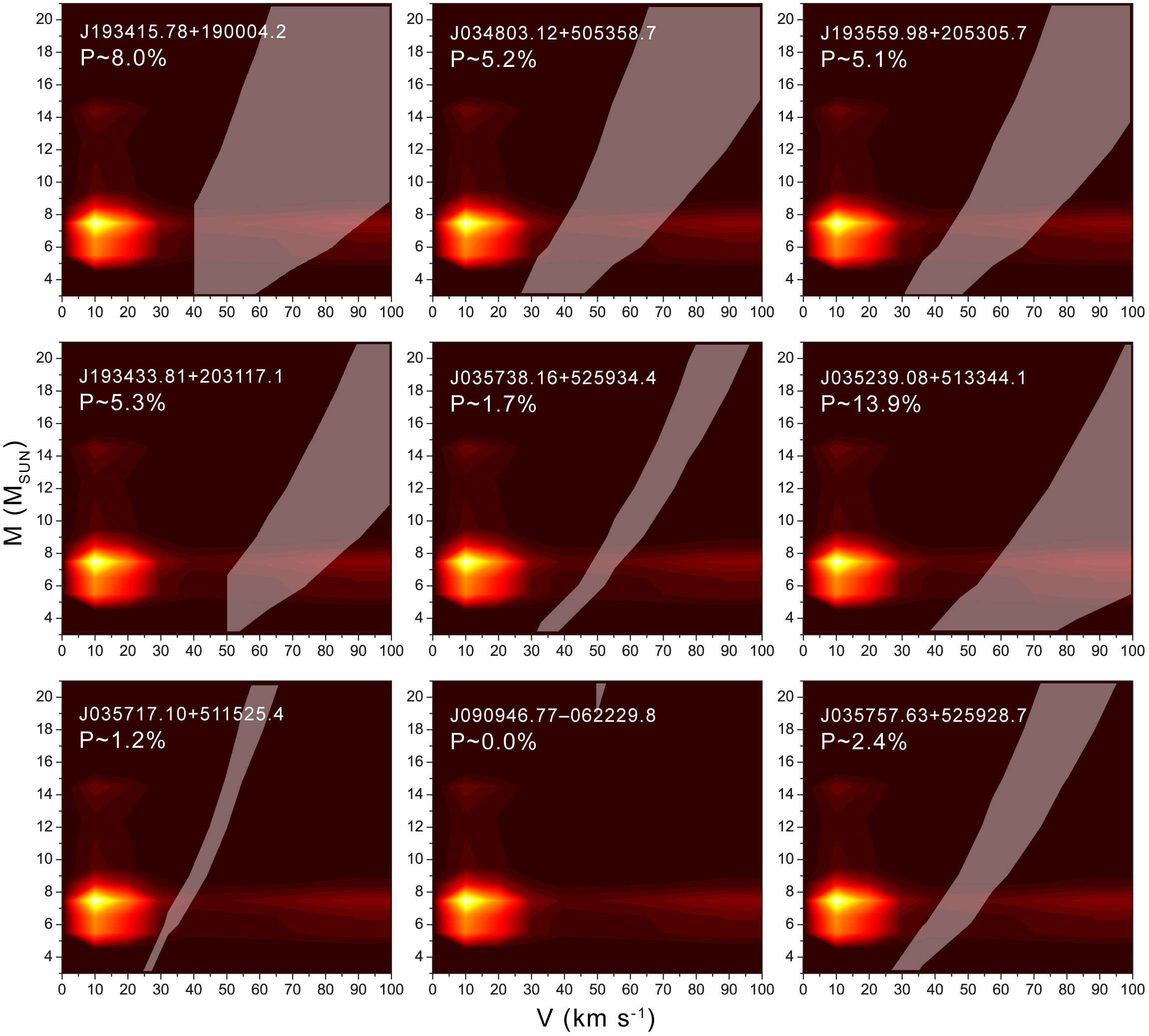}
\caption{\scriptsize{Regions (stripes) of masses $M$ and velocities $V$ for nine candidates, for which the observed and model parameters are in agreement. The stripes are limited from the left by the minimal velocity estimates corresponding to the observed transverse velocity $V_{\rm tr}$ of each object. The probabilities that the candidates are indeed black holes are given in the top left-hand corner of each panel.} }
\label{fig:final}
\end{figure}

This two-dimensional $MV$ distribution was used to estimate the total probability for the selected candidates to be BHs. To this end, the regions obtained above for each of the candidates  (see Fig.~\ref{fig:all-ave}) were superimposed on this distribution. Integrating the two-dimensional $MV$ probability density within the superimposed regions, we derive the total probability $P$ that the object is a BH. For clarity, this procedure is illustrated in Fig.~\ref{fig:final}, which shows the computed regions for the nine candidates combined with the $MV$ plane.

\clearpage
 
Out of the nine selected candidates, eight have shown probabilities $p_i$ ranging from $1.2$\% to
$13.9$\%. One object was excluded from further consideration, since its observed magnitude does not coincide with the theoretical one for any set of parameters. In order to estimate the total probability $P$ for the remaining eight candidates to have at least one BH among them, we can use the expression 
\begin{equation}
P=1-\prod_{i=1}^{8}(1-p_i)\sim 36\%,
\end{equation}
which gives about $36$\%.

\section{Conclusions}
The presented analysis shows that at least one stellar-mass BH may be present among the considered objects with a sufficiently high probability ($\sim36$\%). Establishing this fact would require observations in two directions. The first is multi-band photometry, which would allow one to determine the character of the continuum for these objects and, in the case of its non-thermal nature, would serve as a serious argument in favor of an accretional origin of their emission. Secondly, observations with a high temporal resolution of up to $10^{-5}$--$10^{-6}$~s are needed, which can be used to either detect variations in this range, which are a sign of fragmented accretion onto a BH, or determine the upper limit on the intensity of this emission component. In the case of their detection, the characteristics of such flares would allow one to investigate the region in direct vicinity of the event horizon. We must note that even an absence of the short flares in the case of a non-thermal spectrum, determined by means of multi-band photometry, would not serve as an argument for rejecting the hypothesis of identification of these objects as BHs. This fact alone, for objects in the regions closest to the Sun, is basis enough for a detailed investigation~--- a search for variations, search for polarization, and detailed spectroscopy.

In this work, the search for BHs was carried out in special regions with enhanced probabilities of their localization. The probable pulsar birthplaces  (disruption regions of the binary systems which they were part of) are not the only zones where searching for BHs is feasible. Another type of such regions includes places of high interstellar medium density, where the luminosity of potential BHs would be higher and therefore, they would be easier to detect.

However, this method is not limited to specifically singled out areas and can be expanded to include any point in space to search for isolated stellar mass BHs all over the sky.\\

\section*{Acknowledgements}
This work was supported in part by the grant number 075-
15-2022-262 of the Ministry of science and higher education
of the Russian Federation (13.MNPMU.21.0003). The authors are grateful to the anonymous referee whose comments improved the paper, and to A.~V.~Biryukov for discussion of the work.

\bibliographystyle{plain}
\bibliography{references.bib}

\begin{thebibliography}{10}

\bibitem{2016ApJ...818L..22A}
B.~P. {Abbott}, R.~{Abbott}, T.~D. {Abbott}, M.~R. {Abernathy}, F.~{Acernese},
  K.~{Ackley}, C.~{Adams}, T.~{Adams}, P.~{Addesso}, and {et al.}
\newblock {Astrophysical Implications of the Binary Black-hole Merger
  GW150914}.
\newblock {\em \apjl}, 818(2):L22, February 2016.

\bibitem{2002MNRAS.334..553A}
Eric {Agol} and Marc {Kamionkowski}.
\newblock {X-rays from isolated black holes in the Milky Way}.
\newblock {\em \mnras}, 334(3):553--562, Aug 2002.

\bibitem{2002ApJ...568..289A}
Z.~{Arzoumanian}, D.~F. {Chernoff}, and J.~M. {Cordes}.
\newblock {The Velocity Distribution of Isolated Radio Pulsars}.
\newblock {\em \apj}, 568(1):289--301, March 2002.

\bibitem{2012MNRAS.427..589B}
M.~V. {Barkov}, D.~V. {Khangulyan}, and S.~B. {Popov}.
\newblock {Jets and gamma-ray emission from isolated accreting black holes}.
\newblock {\em \mnras}, 427(1):589--594, Nov 2012.

\bibitem{1967ARA&A...5...25B}
Alan~H. {Batten}.
\newblock {On the Interpretation of Statistics of Double Stars}.
\newblock {\em \araa}, 5:25, Jan 1967.

\bibitem{2008AdSpR..42..523B}
G.~{Beskin}, A.~{Biryukov}, S.~{Karpov}, V.~{Plokhotnichenko}, and V.~{Debur}.
\newblock {Observational appearances of isolated stellar-mass black hole
  accretion Theory and observations}.
\newblock {\em Advances in Space Research}, 42(3):523--532, August 2008.

\bibitem{2005A&A...440..223B}
G.~M. {Beskin} and S.~V. {Karpov}.
\newblock {Low-rate accretion onto isolated stellar-mass black holes}.
\newblock {\em \aap}, 440(1):223--238, Sep 2005.

\bibitem{1991BSAO...31...33B}
G.~M. {Beskin} and S.~N. {Mitronova}.
\newblock {Updated catalogue of DC dwarfs (1987 variant).}
\newblock {\em Bulletin of the Special Astrophysics Observatory}, 31:33--76,
  January 1991.

\bibitem{1998ApJ...506..780B}
Hans~A. {Bethe} and G.~E. {Brown}.
\newblock {Evolution of Binary Compact Objects That Merge}.
\newblock {\em \apj}, 506(2):780--789, Oct 1998.

\bibitem{1974Ap&SS..28...45B}
G.~S. {Bisnovatyi-Kogan} and A.~A. {Ruzmaikin}.
\newblock {The Accretion of Matter by a Collapsing Star in the Presence of a
  Magnetic Field}.
\newblock {\em \apss}, 28(1):45--59, May 1974.

\bibitem{1961BAN....15..265B}
A.~{Blaauw}.
\newblock {On the origin of the O- and B-type stars with high velocities (the
  ``run-away'' stars), and some related problems}.
\newblock {\em \bain}, 15:265, May 1961.

\bibitem{1981sss..book..265B}
N.~G. {Bochkarev}.
\newblock {\em {The interstellar medium and star formation}}, pages 265--325.
\newblock 1981.

\bibitem{1952MNRAS.112..195B}
H.~{Bondi}.
\newblock {On spherically symmetrical accretion}.
\newblock {\em \mnras}, 112:195, Jan 1952.

\bibitem{1944MNRAS.104..273B}
H.~{Bondi} and F.~{Hoyle}.
\newblock {On the mechanism of accretion by stars}.
\newblock {\em \mnras}, 104:273, January 1944.

\bibitem{2000A&AS..143...33B}
F.~{Bonnarel}, P.~{Fernique}, O.~{Bienaym{\'e}}, D.~{Egret}, F.~{Genova},
  M.~{Louys}, F.~{Ochsenbein}, M.~{Wenger}, and J.~G. {Bartlett}.
\newblock {The ALADIN interactive sky atlas. A reference tool for
  identification of astronomical sources}.
\newblock {\em \aaps}, 143:33--40, April 2000.

\bibitem{1987AJ.....94..666C}
R.~G. {Carlberg} and K.~A. {Innanen}.
\newblock {Galactic Chaos and the Circular Velocity at the Sun}.
\newblock {\em \aj}, 94:666, Sep 1987.

\bibitem{2010AstL...36..116C}
E.~G. {Chmyreva}, G.~M. {Beskin}, and A.~V. {Biryukov}.
\newblock {Search for pairs of isolated radio pulsars{\textemdash}Components in
  disrupted binary systems}.
\newblock {\em Astronomy Letters}, 36(2):116--133, Feb 2010.

\bibitem{2005AJ....129.2542C}
Matthew~J. {Collinge}, Michael~A. {Strauss}, Patrick~B. {Hall}, {\v{Z}}eljko
  {Ivezi{\'c}}, Jeffrey~A. {Munn}, David~J. {Schlegel}, Nadia~L. {Zakamska},
  Scott~F. {Anderson}, Hugh~C. {Harris}, Gordon~T. {Richards}, Donald~P.
  {Schneider}, Wolfgang {Voges}, Donald~G. {York}, Bruce {Margon}, and
  J.~{Brinkmann}.
\newblock {Optically Identified BL Lacertae Objects from the Sloan Digital Sky
  Survey}.
\newblock {\em \aj}, 129(6):2542--2561, Jun 2005.

\bibitem{2002astro.ph..7156C}
J.~M. {Cordes} and T.~J.~W. {Lazio}.
\newblock {NE2001.I. A New Model for the Galactic Distribution of Free
  Electrons and its Fluctuations}.
\newblock {\em arXiv e-prints}, pages astro--ph/0207156, July 2002.

\bibitem{1987ApJ...321..780D}
R.~J. {Dewey} and J.~M. {Cordes}.
\newblock {Monte Carlo Simulations of Radio Pulsars and Their Progenitors}.
\newblock {\em \apj}, 321:780, Oct 1987.

\bibitem{1991A&A...248..485D}
A.~{Duquennoy} and M.~{Mayor}.
\newblock {Multiplicity among solar-type stars in the solar neighbourhood. II -
  Distribution of the orbital elements in an unbiased sample.}
\newblock {\em \aap}, 500:337--376, Aug 1991.

\bibitem{2019ApJ...875L...1E}
{Event Horizon Telescope Collaboration}.
\newblock {First M87 Event Horizon Telescope Results. I. The Shadow of the
  Supermassive Black Hole}.
\newblock {\em \apjl}, 875(1):L1, April 2019.

\bibitem{2006ApJ...643..332F}
Claude-Andr{\'e} {Faucher-Gigu{\`e}re} and Victoria~M. {Kaspi}.
\newblock {Birth and Evolution of Isolated Radio Pulsars}.
\newblock {\em \apj}, 643(1):332--355, May 2006.

\bibitem{2013MNRAS.430.1538F}
R.~P. {Fender}, T.~J. {Maccarone}, and I.~{Heywood}.
\newblock {The closest black holes}.
\newblock {\em \mnras}, 430(3):1538--1547, Apr 2013.

\bibitem{2018A&A...616A...1G}
{Gaia Collaboration}, A.~G.~A. {Brown}, A.~{Vallenari}, T.~{Prusti}, J.~H.~J.
  {de Bruijne}, C.~{Babusiaux}, C.~A.~L. {Bailer-Jones}, M.~{Biermann}, D.~W.
  {Evans}, L.~{Eyer}, and {et al}.
\newblock {Gaia Data Release 2. Summary of the contents and survey properties}.
\newblock {\em \aap}, 616:A1, Aug 2018.

\bibitem{2011MNRAS.417.1210G}
J.~{Girven}, B.~T. {G{\"a}nsicke}, D.~{Steeghs}, and D.~{Koester}.
\newblock {DA white dwarfs in Sloan Digital Sky Survey Data Release 7 and a
  search for infrared excess emission}.
\newblock {\em \mnras}, 417(2):1210--1235, Oct 2011.

\bibitem{1970ApJ...160L..91G}
III {Gott}, J.~Richard, James~E. {Gunn}, and Jeremiah~P. {Ostriker}.
\newblock {Runaway Stars and the Pulsars Near the Crab Nebula}.
\newblock {\em \apjl}, 160:L91, May 1970.

\bibitem{2019ApJ...887...93G}
Gregory~M. {Green}, Edward {Schlafly}, Catherine {Zucker}, Joshua~S. {Speagle},
  and Douglas {Finkbeiner}.
\newblock {A 3D Dust Map Based on Gaia, Pan-STARRS 1, and 2MASS}.
\newblock {\em \apj}, 887(1):93, December 2019.

\bibitem{2009MNRAS.400.2050G}
Tolga {G{\"u}ver} and Feryal {{\"O}zel}.
\newblock {The relation between optical extinction and hydrogen column density
  in the Galaxy}.
\newblock {\em \mnras}, 400(4):2050--2053, December 2009.

\bibitem{2003A&A...397..159H}
J.~L. {Halbwachs}, M.~{Mayor}, S.~{Udry}, and F.~{Arenou}.
\newblock {Multiplicity among solar-type stars. III. Statistical properties of
  the F7-K binaries with periods up to 10 years}.
\newblock {\em \aap}, 397:159--175, Jan 2003.

\bibitem{1977ApJ...216..842H}
D.~J. {Helfand} and E.~{Tademaru}.
\newblock {Pulsar velocity observations: correlations, interpretations, and
  discussion.}
\newblock {\em \apj}, 216:842--851, Sep 1977.

\bibitem{2005MNRAS.360..974H}
G.~{Hobbs}, D.~R. {Lorimer}, A.~G. {Lyne}, and M.~{Kramer}.
\newblock {A statistical study of 233 pulsar proper motions}.
\newblock {\em \mnras}, 360(3):974--992, Jul 2005.

\bibitem{1996ApJ...456..738I}
Jr. {Iben}, Icko and Alexander~V. {Tutukov}.
\newblock {On the Origin of the High Space Velocities of Radio Pulsars}.
\newblock {\em \apj}, 456:738, Jan 1996.

\bibitem{1982ApJ...255..654I}
J.~R. {Ipser} and R.~H. {Price}.
\newblock {Synchrotron radiation from spherically accreting black holes}.
\newblock {\em \apj}, 255:654--673, Apr 1982.

\bibitem{2019MNRAS.489.2038I}
P.~B. {Ivanov}, V.~N. {Lukash}, S.~V. {Pilipenko}, and M.~S. {Pshirkov}.
\newblock {Search for isolated Galactic Centre stellar mass black holes in the
  IR and sub-mm range}.
\newblock {\em \mnras}, 489(2):2038--2048, Oct 2019.

\bibitem{1989MNRAS.239..571K}
Konrad {Kuijken} and Gerard {Gilmore}.
\newblock {The mass distribution in the galactic disc. I - A technique to
  determine the integral surface mass density of the disc near the sun.}
\newblock {\em \mnras}, 239:571--603, Aug 1989.

\bibitem{2018MNRAS.480.2704L}
A.~{Lamberts}, S.~{Garrison-Kimmel}, P.~F. {Hopkins}, E.~{Quataert}, J.~S.
  {Bullock}, C.~A. {Faucher-Gigu{\`e}re}, A.~{Wetzel}, D.~{Kere{\v{s}}},
  K.~{Drango}, and R.~E. {Sanderson}.
\newblock {Predicting the binary black hole population of the Milky Way with
  cosmological simulations}.
\newblock {\em \mnras}, 480(2):2704--2718, October 2018.

\bibitem{2016ApJ...830...41L}
J.~R. {Lu}, E.~{Sinukoff}, E.~O. {Ofek}, A.~{Udalski}, and S.~{Kozlowski}.
\newblock {A Search For Stellar-mass Black Holes Via Astrometric Microlensing}.
\newblock {\em \apj}, 830(1):41, Oct 2016.

\bibitem{2005MNRAS.360L..30M}
Thomas~J. {Maccarone}.
\newblock {Using radio emission to detect isolated and quiescent accreting
  black holes}.
\newblock {\em \mnras}, 360(1):L30--L34, Jun 2005.

\bibitem{2005yCat.7245....0M}
R.~N. {Manchester}, G.~B. {Hobbs}, A.~{Teoh}, and M.~{Hobbs}.
\newblock {VizieR Online Data Catalog: ATNF Pulsar Catalog (Manchester+,
  2005)}.
\newblock {\em VizieR Online Data Catalog}, page VII/245, August 2005.

\bibitem{1975A&A....44...59M}
P.~{Meszaros}.
\newblock {Radiation from spherical accretion onto black holes.}
\newblock {\em \aap}, 44(1):59--68, Nov 1975.

\bibitem{1981gask.book.....M}
D.~{Mihalas} and J.~{Binney}.
\newblock {\em {Galactic astronomy. Structure and kinematics}}.
\newblock 1981.

\bibitem{2003ApJ...594..936P}
Rosalba {Perna}, Ramesh {Narayan}, George {Rybicki}, Luigi {Stella}, and Aldo
  {Treves}.
\newblock {Bondi Accretion and the Problem of the Missing Isolated Neutron
  Stars}.
\newblock {\em \apj}, 594(2):936--942, September 2003.

\bibitem{2010AJ....139..390P}
Richard~M. {Plotkin}, Scott~F. {Anderson}, W.~N. {Brandt}, Aleksandar~M.
  {Diamond-Stanic}, Xiaohui {Fan}, Patrick~B. {Hall}, Amy~E. {Kimball},
  Michael~W. {Richmond}, Donald~P. {Schneider}, Ohad {Shemmer}, Wolfgang
  {Voges}, Donald~G. {York}, Neta~A. {Bahcall}, Stephanie {Snedden}, Dmitry
  {Bizyaev}, Howard {Brewington}, Viktor {Malanushenko}, Elena {Malanushenko},
  Dan {Oravetz}, Kaike {Pan}, and Audrey {Simmons}.
\newblock {Optically Selected BL Lacertae Candidates from the Sloan Digital Sky
  Survey Data Release Seven}.
\newblock {\em \aj}, 139(2):390--414, February 2010.

\bibitem{1998A&A...332..173P}
S.~F. {Portegies Zwart} and L.~R. {Yungelson}.
\newblock {Formation and evolution of binary neutron stars}.
\newblock {\em \aap}, 332:173--188, Apr 1998.

\bibitem{2021MNRAS.505.4036S}
Francesca {Scarcella}, Daniele {Gaggero}, Riley {Connors}, Julien {Manshanden},
  Massimo {Ricotti}, and Gianfranco {Bertone}.
\newblock {Multiwavelength detectability of isolated black holes in the Milky
  Way}.
\newblock {\em \mnras}, 505(3):4036--4047, August 2021.

\bibitem{1969AZh....46..715S}
I.~S. {Shklovskii}.
\newblock {Possible Causes of the Secular Increase in Pulsar Periods.}
\newblock {\em \azh}, 46:715, Jan 1969.

\bibitem{1971SvA....15..377S}
V.~F. {Shvartsman}.
\newblock {Halos around ``Black Holes''.}
\newblock {\em \sovast}, 15:377, Dec 1971.

\bibitem{1977SoSAO..19....5S}
V.~F. {Shvartsman}.
\newblock {The MANIA [Multichannel Analysis of Nanosecond Intensity
  Alterations] experiment. Astrophysical problems, mathematical methods,
  instrumentation complex, results of the first observations.}
\newblock {\em Soobshcheniya Spetsial'noj Astrofizicheskoj Observatorii},
  19:5--38, Jan 1977.

\bibitem{1989SvAL...15..145S}
V.~F. {Shvartsman}, G.~M. {Beskin}, and S.~N. {Mitronova}.
\newblock {A Search for 0.5-MICROSECOND to 40-SECOND Optical Variability in DC
  White Dwarfs}.
\newblock {\em Soviet Astronomy Letters}, 15:145, Apr 1989.

\bibitem{1989Afz....31..457S}
V.~F. {Shvartsman}, G.~M. {Beskin}, and S.~A. {Pustil'Nik}.
\newblock {The results of search for superrapid optical variability of radio
  objects with continuous optical spectra.}
\newblock {\em Astrofizika}, 31:457--465, Jan 1989.

\bibitem{2019MNRAS.488.2099T}
Daichi {Tsuna} and Norita {Kawanaka}.
\newblock {Radio emission from accreting isolated black holes in our galaxy}.
\newblock {\em \mnras}, 488(2):2099--2107, Sep 2019.

\bibitem{2018MNRAS.477..791T}
Daichi {Tsuna}, Norita {Kawanaka}, and Tomonori {Totani}.
\newblock {X-ray detectability of accreting isolated black holes in our
  Galaxy}.
\newblock {\em \mnras}, 477(1):791--801, Jun 2018.

\bibitem{2004ApJ...610..402V}
W.~H.~T. {Vlemmings}, J.~M. {Cordes}, and S.~{Chatterjee}.
\newblock {Separated at Birth: The Origin of the Pulsars B2020+28 and B2021+51
  in the Cygnus Superbubble}.
\newblock {\em \apj}, 610(1):402--410, July 2004.

\bibitem{2000A&AS..143....9W}
M.~{Wenger}, F.~{Ochsenbein}, D.~{Egret}, P.~{Dubois}, F.~{Bonnarel},
  S.~{Borde}, F.~{Genova}, G.~{Jasniewicz}, S.~{Lalo{\"e}}, S.~{Lesteven}, and
  R.~{Monier}.
\newblock {The SIMBAD astronomical database. The CDS reference database for
  astronomical objects}.
\newblock {\em \aaps}, 143:9--22, April 2000.

\bibitem{2019ApJ...885....1W}
Grzegorz {Wiktorowicz}, {\L}ukasz {Wyrzykowski}, Martyna {Chruslinska}, Jakub
  {Klencki}, Krzysztof~A. {Rybicki}, and Krzysztof {Belczynski}.
\newblock {Populations of Stellar-mass Black Holes from Binary Systems}.
\newblock {\em \apj}, 885(1):1, November 2019.

\end{thebibliography}
 
\end{document}